\documentclass[floatfix,twocolumn,rmp,showpacs,superscriptaddress]{revtex4}

\usepackage{dcolumn,graphicx,amsmath,amssymb,txfonts}

\begin{document}

\title{Efficient local strategies for vaccination and network attack}

\author{Petter Holme}
\email{holme@tp.umu.se}
\affiliation{Department of Physics, Ume{\aa} University, 901 87
  Ume{\aa}, Sweden}
\affiliation{NORDITA, Blegdamsvej 17, 2100 Copenhagen, Denmark}

\begin{abstract} % paper --> Letter (for APS)
  We study how a fraction of a population should be vaccinated to most
  efficiently stop epidemics. We argue that only local information
  (about the neighborhood of specific vertices) is usable in practice,
  and hence we consider only local vaccination strategies. The
  efficiency of the vaccination strategies is investigated with both
  static and dynamical measures. Among other things we find that the
  most efficient strategy for many real-world situations is to
  iteratively vaccinate the neighbor of the previous vaccinee that has
  most links out of the neighborhood.
\end{abstract}

\pacs{89.65.--s, 89.75.Hc, 89.75.--k}

\maketitle

\section{Introduction}

Diseases spread over networks. The spreading dynamics are closely
related to the structure of networks. For this reason network
epidemiology has turned into of the most vibrant subdisciplines of
complex network studies.~\cite{gies,lea:sex,mejn:rev} A topic of great
practical importance within network epidemiology is the vaccination
problem: How should a population be vaccinated to most efficiently
prevent a disease to turn into an epidemic? For economic reasons it is
often not possible to vaccinate the whole population. Some vaccines
have severe side effects and for this reason one may also want to keep
number of vaccinated individuals low. So if cheap vaccines, free of
side effects, does not exist; then having an efficient vaccination
strategy is essential for saving both money and life. If all ties
within the population is known, then the target persons for
vaccination can be identified using sophisticated global strategies
(cf.~\cite{our:attack}); but this is hardly possible for nation-wide
(or larger) vaccination campaigns. In a seminal paper Cohen \textit{et
  al.}~\cite{chn:vacc} suggested a vaccination strategy that only
requires a person to estimate which other persons he, or she, gets
close enough to for the disease to spread to---i.e., to name the
``neighbors'' in the network over which the disease spreads. For
network with a skewed distribution of degree (number of neighbors) the
strategy to vaccinate a neighbor of a randomly chosen person is much
more efficient than a random vaccination. In this work we assume that each
individual knows a little bit more about his, or her, neighborhood
than just the names of the neighbors: We also assume that an
individual can guess the degree of the neighbors and the ties from one
neighbor to another. This assumption is not very unrealistic---people
are believed to have a good understanding of their social
surroundings (this is, for example, part of the explanation for the
``navigability'' of social networks)~\cite{watts:search}. 

Finding the optimal set of vaccinees is closely related to the attack
vulnerability problem~\cite{our:attack,alb:attack}. The
major difference is the dynamic system that is confined to the
network---disease spreading for the vaccination problem and
information flow for the attack vulnerability problem. To be able to
protect the network efficiently one needs to know the worst case
attacking scenario. Large scale network attacks are, presumably, based
on local (rather than global) network information. So, a
grave scenario would be in the network was attacked with the same
strategy that is most efficient for vaccination. We will use the
vaccination problem as the framework for our discussion, but the
results applies for network attack as well.

\section{Preliminaries}

In our discussion we will use two measures of network structure: The
\textit{clustering coefficient} $C$ of the network defined as the
ratio of triangles with respect to connected triples normalized to the
interval $[0,1]$.~\cite{bw:sw} If $C=1$ there is a maximal number of
triangles (given a set of connected triples); if $C=0$ the graph has
no triangles. We also measure the degree-degree correlations through
the \textit{assortative mixing
  coefficient} defined as~\cite{mejn:assmix}
\begin{equation}
  r=\frac{4\langle k_1\, k_2\rangle - \langle k_1 + k_2\rangle^2}
  {2\langle k_1^2+k_2^2\rangle - \langle k_1+ k_2\rangle^2}~,
\end{equation}
where $k_i$ is the degree of the $i$'th argument of an edge in a list
of the edges, and $\langle\:\cdot\:\rangle$ denotes average over
that edge-list. We let $N$ denote the number of
vertices and $M$ the number of edges.

\section{The networks}

We will test the vaccination strategies we propose on both real-world
and model networks.  
The first real-world network is a scientific
collaboration network~\cite{mejn:scicolpnas}. The vertices of this
network are scientists who have uploaded manuscripts to the preprint
repository arxiv.org. An edge between two authors means that
they have coauthored a preprint. We also study two small real-world
social networks: One constructed from an observational study of
friendships in a karate club, another based on an interview survey
among prisoners. The edges of these small networks are, probably, more
relevant for disease spreading than the arxiv network, but may suffer
from finite size effects. The three model networks are: 1. The Holme-Kim
(HK) model~\cite{hk:model} that produces networks with a power-law degree
distribution and tunable clustering. Basically, it is a
Barab\'{a}si-Albert (BA) type growth model based on preferential
attachment~\cite{ba:model}---just as the BA model
it has one parameter $m=M/N$ controlling the average degree and one
(additional) parameter $m_t\in [1,m]$ controlling the clustering. We
will use $M=2N=4000$ and $m=m_t+1=4$ giving the maximal clustering for
the given $N$ and $M$. 2. The networked seceder model, modeling social
networks with a community structure and exponentially decaying
degree distributions~\cite{our:seceder}. Briefly, it works by
sequentially updating the vertices by, for each vertex $v$, rewiring
all $v$'s edges to the neighborhood of a peripheral vertex. With a
probability $r$ an edge of $v$ can be rewired to a random vertex (so
$r$ controls the degree of community structure). We use the parameter
values $M=3N=6600$, $r=0.1$ and $10M$ iterations on an
Erd\H{o}s-R\'{e}nyi network~\cite{er:on}. 3. The Watts-Strogatz (WS)
model~\cite{wattsstrogatz} generates networks with exponentially decaying
degree distributions and tunable clustering. The WS model starts from
the vertices on a circular topology with edges between vertices
separated by 1 to $k$ steps on the circle. Then one goes through the
edges and rewire one side of them to randomly selected vertices with a
probability $P$. We use $P=0.05$ and $M=kN=2N=4000$.

\begin{table}
\caption{Statistics of the networks. Note that the arxiv, prison and
  seceder model networks are not connected---the largest connected
  components contains $48561$, $58$ and $2162(1)$ nodes respectively.
}
\label{tab:stat}
\begin{ruledtabular}
  \begin{tabular}{l|llll}
    network & $N$ & $M$ & $C$ & $r$ \\\hline
    arxiv & 58342 & 294901 & 0.420 & +0.324 \\
    karate club & 34 & 78 & 0.256 & --0.476\\
    prison & 67 & 85 & 0.310 & +0.161\\
    HK & 2000 & 4000 & 0.1753(1) & --0.0458(1) \\
    seceder & 2200 & 6600 & 0.266(1) & +0.012(2)\\
    WS & 2000 & 4000 & 0.4219(1) & --0.01267(2) \\
  \end{tabular}
\end{ruledtabular}
\end{table}

\begin{figure*}
  \resizebox*{\linewidth}{!}{\includegraphics{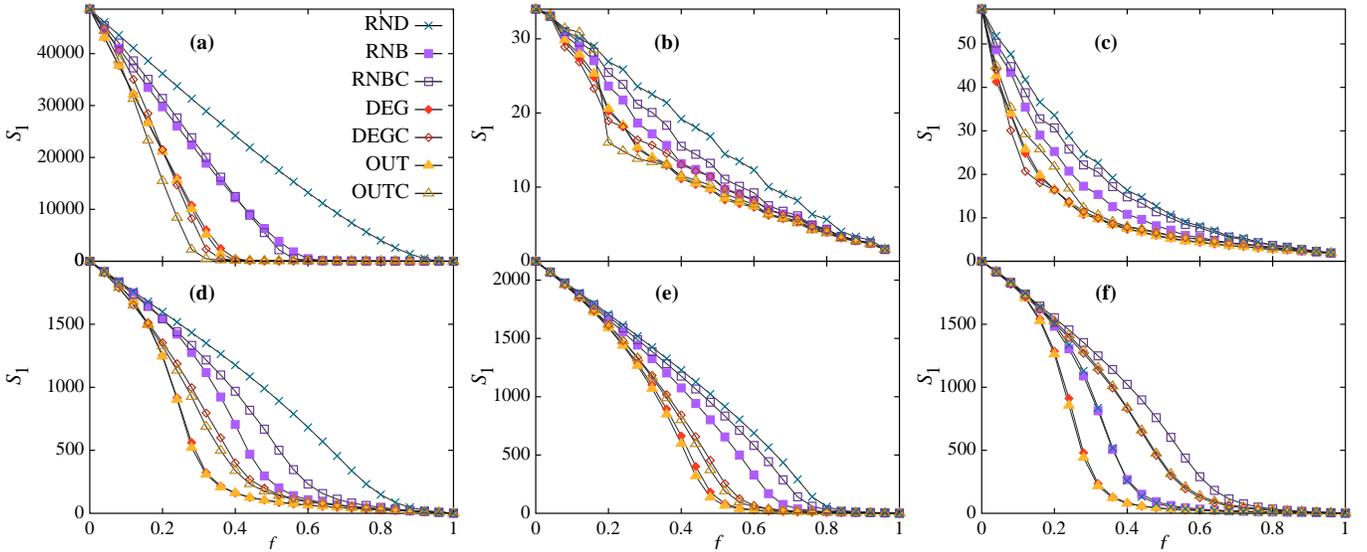}}
  \caption{
    The size of the largest connected component $S_1$ as a function of
  the fraction of vaccinated vertices for the (a) arxiv, (b) karate
  club, (c) prison, (d) HK model, (e) seceder model and (f) WS model
  network. Error bars are smaller than the symbol size. Lines are
  guides for the eyes.
  }
  \label{fig:s1}
\end{figure*}

\section{The strategies}

Now we turn to the definition of the strategies. We assume a fraction
$f$ of the population is to be vaccinated. As a reference we consider
random vaccination (\textsc{Rnd}, equivalent to site percolation). We use
the above mentioned \textit{neighbor vaccination} (\textsc{RNb})---to
vaccinate the neighbor of randomly chosen vertices---and the trivial improvement~\cite{bjk:pfs} if
knowledge about the neighbors' degrees are included: Pick a vertex at
random and vaccinate one (randomly chosen) of its highest-degree
neighbors (we call it \textsc{Deg}). To avoid overvaccination of a
neighborhood one can consider to vaccinate neighbors of a vertex $v$
with a maximal number of edges out of $v$'s neighborhood
(\textsc{Out}). For all strategies except \textsc{Rnd}
we also consider ``chained'' versions were one, instead of vaccinating a
neighbor of a randomly chosen vertex, vaccinates a neighbor of the vertex
vaccinated in the previous time step (if all neighbors are vaccinated
a neighbor of a random vertex is chosen instead). For the acronyms of
the chained versions a suffix ``C'' is added.

\begin{figure}
  \resizebox*{0.9 \linewidth}{!}{\includegraphics{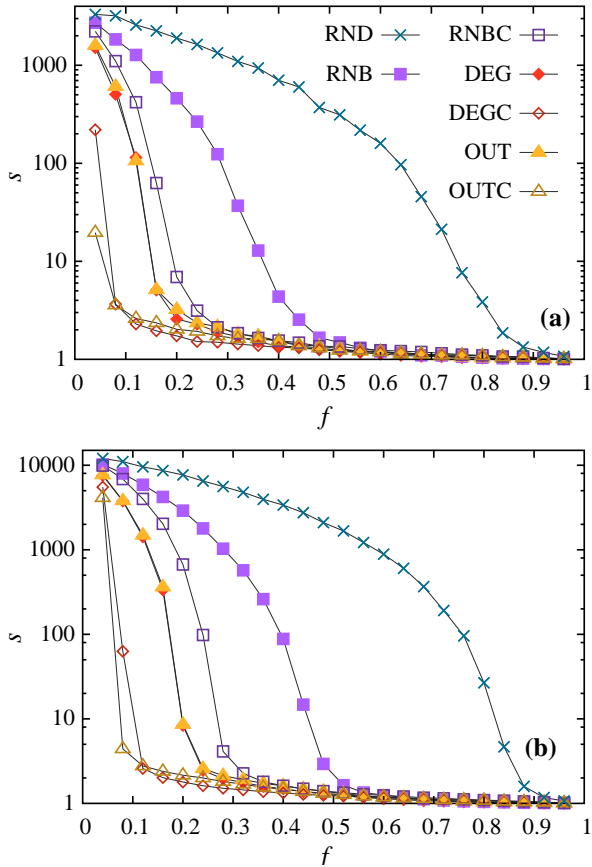}}
  \caption{The average number of vertices that are at infected once or
  more during an outbreak $s$ for (a) the SIR and (b) the SIS disease
  dynamics. Error bars of the order of the symbol size. Lines are
  guides for the eyes.}
  \label{fig:dyn}
\end{figure}

\section{Results and analysis}

The results of this paper are presented in three sections: First we
study how the number of vertices in the largest connected subgraph
$S_1$ depends on the fraction $f$ of vaccinated vertices. Then we
show that the conclusions from $S_1$ also hold for dynamical simulations
of disease spreading. To interpret the results we also investigate
$S_1$ for a fixed $f$ as a function of the clustering and assortative
mixing coefficients.

\subsection{Static efficiency}

As a static efficiency measure we consider the size of the average
largest connected component of susceptible (non-vaccinated) vertices,
$S_1$. We average over $n_\mathrm{vac}=1000$ runs of the vaccination
procedures. The model networks are also averaged over
$n_\mathrm{net}=100$ network realizations. (Smaller or larger
$n_\mathrm{vac}$ and $n_\mathrm{net}$ does not make any qualitative
difference.) In Fig.~\ref{fig:s1} we display $S_1$ as a
function of $f$. For all except the WS model network the \textsc{Deg}
and \textsc{Out} (chained and unchained versions) form the most
efficient set of strategies. Within this group the order of efficiency
varies: For the arxiv network the \textsc{Out} strategy is more than
twice as efficient as any other for $0.25\lesssim f\lesssim 0.4$. For
the HK and seceder model networks the chained strategies are
considerably more efficient than the unchained ones. We note that the
difference between the chained and unchained versions of \textsc{Out}
and \textsc{Deg} is bigger than between \textsc{Out} and \textsc{Deg}
(or \textsc{OutC} and \textsc{DegC}). \textsc{Out} do converge to
\textsc{Deg} in the limit of vanishing $C$ but all networks we test
have rather high clustering. Another interesting observation is that
even if the degree distribution is narrow, such as for the seceder
model of Fig.~\ref{fig:s1}(e) (where $P(k)\sim \exp(-k)$) the more
elaborate strategies are much more efficient than random
vaccination. This is especially clear for higher $f$ which suggests
that the structural change of the network of susceptible vertices
during the vaccination procedure is an important factor for the
overall efficiency. For the WS model network the chained algorithms
are performing poorer than random vaccination. This is in contrast to
all other networks. We conclude that epidemiology related results
regarding the WS model networks should be cautiously generalized to
real-world systems.

\subsection{Dynamic efficiency}

Static measures of vaccination efficiency are potential
over-simplifications---there is a chance that the interplay between
disease dynamics and the underlying network structure has a
significant role. To motivate the use of $S_1$ we also investigate the
SIS and SIR models~\cite{gies} on vaccinated networks. In the SIS
model a vertex goes from ``susceptible'' (S) to ``infected'' (I) and
back to S. In the SIR model is just the same, except that an
infected vertex goes to the ``removed'' (R) state and remain
there. The probability to go from $S$ to $I$ (per contact) is zero for
vaccinated vertices and $\lambda=0.05$ for the rest. The I state lasts
$\delta=2.5$ time steps. We use synchronous updating and one randomly
chosen initially infected
person. The disease dynamics are averaged $n_\mathrm{dis}=100$ times for
all $n_\mathrm{vac}=1000$ runs of the vaccination schemes.  In
Fig.~\ref{fig:dyn}(a) we plot the average number of individuals that
at least once have been infected during an outbreak $s$---i.e., until
there are no I-vertices left, or (for SIS) has reached an endemic
state (defined in the simulations as when there are no susceptible
vertices that have not had the disease at least once)---for the arxiv
network. Other networks and simulation parameters give qualitatively
similar results. Qualitatively, the large picture from the $S_1$
calculations remains---the chained and unchained \textsc{Deg} and
\textsc{Out} strategies are very efficient, and the chained versions
are more efficient than the unchained. A difference is that the
unchained \textsc{RNb} also performs rather well. Quantitatively, the
differences between the strategies are huge, this is a result of the
threshold behaviors of the SIS and SIR models~\cite{chn:vacc}. The
conclusion of Fig.~\ref{fig:dyn} (and similar plots for other
networks) is that the order of the strategies' efficiencies are
largely the same as concluded from the $S_1(f)$-curves. But if high
resolution is required, the measurement of network fragility has to be
specific for the studied system.

\begin{figure}
  \resizebox*{0.9 \linewidth}{!}{\includegraphics{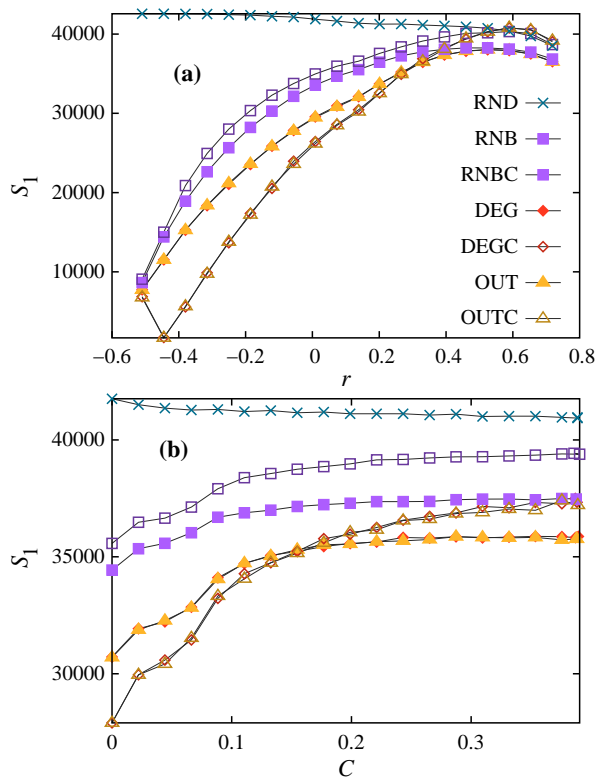}}
  \caption{How the size of the largest connected component vaccination
    of 20\% of the population depends on clustering and degree-degree
  correlations. (a) shows $S_1(f=0.2)$ plotted against $r$. (b) shows
  $S_1(f=0.2)$
  as a function of $C$. The networks have the same size and degree
  sequence as the arxiv network. Error bars are smaller than the
  symbol size. Lines are guides for the eyes.}
  \label{fig:rew}
\end{figure}

\subsection{The role of clustering and assortative mixing}

To gain some insight how the network structure govern the relative
efficiencies of the strategies we measure $S_1(f=0.2)$ for varying
assortative mixing and clustering coefficients. The results hold for
other small $f$ values. We keep the size and
degree sequence constant to the values of the arxiv network. To
perform this sampling we rewire pairs of edges $(v_1,v_2)$ and
$(w_1,w_2)$ to $(v_1,w_2)$ and $(w_1,v_2)$ (unless this would
introduce a self-edge or multiple edges). To ensure that the
$n_\mathrm{rew}=100$ rewiring realizations are independent we start
with rewiring $n_\mathrm{init}=3M$ pairs of edges. Then we go through
pairs of edges randomly and execute only changes that makes the
current $r$ or $C$ closer to their target values. When the value of
$r$ or $C$ are within $0.1\%$ of the target value the iteration is
braked. The results seen in Fig.~\ref{fig:rew} shows that, just as
before the \textsc{Out} and \textsc{Deg} strategies, chained or
unchained, are most efficient throughout the parameter space. The
unchained versions are most efficient for $r\gtrsim 0.3$. An
explanation is that, for high $r$, the chained versions will
effectively only vaccinate the
high-connected vertices (that are grouped together for very high $r$)
and leave chains of low-degree vertices unvaccinated. The
$C$-dependence plotted in Fig.~\ref{fig:rew}(b) shows that the
unchained versions outperform the chained versions for $C\gtrsim
0.15$. This is possibly a result of that the chains, for
combinatorial reasons, get stuck in one part of the network. It is not
an effect of biased
degree-degree correlations since if the rewiring procedure is conditioned to a
fixed $r$ Fig.~\ref{fig:rew}(b) remains essentially unaltered. We note
that the structure of the original arxiv network differs from the
rewired networks. For example, at $f=0.2$ of Fig.~\ref{fig:s1}(a) the
\textsc{Out} is 22\% more efficient than \textsc{OutC}, but in
Fig.~\ref{fig:rew} the \textsc{Out} and \textsc{OutC} curves differ
very little. For the \textsc{RNb} strategy the chained version is better than
the unchained throughout the range of $r$ and $C$ values.

\section{Summary and conclusions}

To summarize, we have investigated strategies for vaccination and
network attack that are based only on the knowledge of the
neighborhood---information that humans arguably possess and
utilize. Both static and dynamical measures of efficiency are
studied. For most networks, regardless of the number of vaccinated
vertices, the most efficient strategies are to choose a vertex $v$ and
vaccinate a neighbor of $v$ with highest degree (\textsc{Deg}), or the
neighbor of $v$ with most links out of $v$'s neighborhood
(\textsc{Out}). $v$ can be picked either as the lastly vaccinated
vertex (chained selection) or at random (unchained selection). For
real-world networks the chained versions tend to outperform the
unchained ones, whereas this situation is reversed for the three types
of model networks we study. We investigate the relative efficiency of
chained and unchained strategies further by sampling random networks
with a fixed degree sequence and varying assortative mixing and
clustering coefficients. We find that the unchained strategies are
preferable for networks with a very high clustering or strong
positive assortative mixing (larger values than in seen in real-world
networks). In Ref.~\cite{chn:vacc} the authors propose
the strategy to vaccinate  a random neighbor of a randomly selected
vertex. This strategy (\textsc{RNb}) requires less information of the
neighborhood than \textsc{Deg} and \textsc{Out} do. Thus the
practical procedure gets simpler: One only has to ask a person
``name a person you meet regularly'' rather than ``name the acquaintance of yours who meet most people you are not
acquainted with regularly'' (for \textsc{Out}). (``Meet with regularly''
should be replaced with some phrase signifying a high risk of infection
transfer for the pathogen in question.) On the other hand, if the
information of the neighborhoods is incomplete \textsc{Deg} and
\textsc{Out} will, effectively, be reduced to \textsc{RNb} (and thus not
perform worse than \textsc{RNb}). To epitomize, choosing the people to
vaccinate in the right way will save a tremendous amount of vaccine
and side-effect cases. The best strategy can only be selected by
considering both the structure of the network the pathogen spreads
over, and the disease dynamics. If nothing of this is known the
\textsc{OutC} strategy our recommendation---it is better, or not much
worse, than the best strategy in most cases. Together with
\textsc{DegC}, \textsc{OutC} is most efficient for low clustering
and assortative mixing coefficients, which is the region of parameter
space for sexually transmitted diseases---the most interesting case
for network based vaccination schemes (due to the well-definedness of
sexual networks).

\section*{Acknowledgements}

The author is grateful for comments from M.\ Rosvall and acknowledges
support from the Swedish Research Council through contract no.\
2002-4135.


\begin{thebibliography}{10}

\bibitem{alb:attack}
R.~Albert, H.~Jeong, and A.-L. Barab\'{a}si, \textit{Attack and error tolerance
  of complex networks}, Nature \textbf{406} (2000), pp.~378-382.

\bibitem{ba:model}
A.-L. Barab\'{a}si and R.~Albert, \textit{Emergence of scaling in random
  networks}, Science \textbf{286} (1999), pp.~509-512.

\bibitem{bw:sw}
A.~Barrat and M.~Weigt, \textit{On the properties of small-world network
  models}, Eur. Phys. J. B \textbf{13} (2000), pp.~547-560.

\bibitem{chn:vacc}
R.~Cohen, S.~Havlin, and D.~ben Avraham, \textit{Efficient immunization
  strategies for computer networks and populations}, Phys. Rev. Lett.
  \textbf{91} (2003), art.~no.\ 247901.

\bibitem{er:on}
P.~Erd\H{o}s and A.~R\'{e}nyi, \textit{On random graphs {I}}, Publ. Math.
  Debrecen \textbf{6} (1959), pp.~290-297.

\bibitem{gies}
J.~Giesecke, \textit{Modern infectious disease epidemiology}, Arnold, London,
  2~ed., 2002.

\bibitem{our:seceder}
A.~Gr\"{o}nlund and P.~Holme, \textit{Networking the seceder model: Group
  formation in social and economic systems}.
\newblock e-print: cond-mat/0312010.

\bibitem{hk:model}
P.~Holme and B.~J. Kim, \textit{Growing scale-free networks with tunable
  clustering}, Phys. Rev. E \textbf{65} (2002), art.~no.\ 026107.

\bibitem{our:attack}
P.~Holme, B.~J. Kim, C.~N. Yoon, and S.~K. Han, \textit{Attack vulnerability of
  complex networks}, Phys. Rev. E \textbf{65} (2002), art.~no.\ 066109.

\bibitem{bjk:pfs}
B.~J. Kim, C.~N. Yoon, S.~K. Han, and H.~Jeong, \textit{Path finding strategies
  in scale-free networks}, Phys. Rev. E \textbf{65} (2002), art.~no.\ 027103.

\bibitem{lea:sex}
F.~Liljeros, C.~R. Edling, and L.~A. {Nunes Amaral}, \textit{Sexual networks:
  Implication for the transmission of sexually transmitted infection}, Microbes
  Infect. \textbf{5} (2003), pp.~189-196.

\bibitem{mejn:scicolpnas}
M.~E.~J. Newman, \textit{The structure of scientific collaboration networks},
  Proc. Natl. Acad. Sci. USA \textbf{98} (2001), pp.~404-409.

\bibitem{mejn:assmix}
\leavevmode\vrule height 2pt depth -1.6pt width 23pt, \textit{Assortative
  mixing in networks}, Phys. Rev. Lett. \textbf{89} (2002), art.~no.\ 208701.

\bibitem{mejn:rev}
\leavevmode\vrule height 2pt depth -1.6pt width 23pt, \textit{The structure and
  function of complex networks}, SIAM Rev. \textbf{45} (2003), pp.~167-256.

\bibitem{watts:search}
D.~J. Watts, P.~S. Dodds, and M.~E.~J. Newman, \textit{Identity and search in
  social networks}, Science \textbf{296} (2002), pp.~1302-1305.

\bibitem{wattsstrogatz}
D.~J. Watts and S.~H. Strogatz, \textit{Collective dynamics of {`small-world'}
  networks}, Nature \textbf{393} (1998), pp.~440-442.

\end{thebibliography}
\end{document}